\documentclass{aa} 
\usepackage{graphicx}
\usepackage{txfonts}
\usepackage{natbib}
\bibpunct{(}{)}{;}{a}{}{,}

\begin{document}
   \title{Cosmological evolution of compact AGN at 15 GHz}


   \author{T.~G.~Arshakian\inst{1}\fnmsep\thanks{On leave from
           Byurakan Astrophysical Observatory, Byurakan 378433,
           Armenia and Isaac Newton Institute of Chile, Armenian
           Branch}, E.~Ros\inst{1} \and J.~A.~Zensus\inst{1} }
           \offprints{T.~G.~Arshakian}

   \institute{Max-Planck-Institut f\"ur Radioastronomie, Auf
              dem H\"ugel 69, 53121 Bonn, Germany  \\
              \email{tigar@mpifr-bonn.mpg.de, ros@mpifr-bonn.mpg.de,
              azensus@mpifr-bonn.mpg.de} }


   \date{Submitted  October 3, 2005}
   
   \abstract
   {}
   {We study the uniformity of the distribution of compact
   flat-spectrum AGN on the sky and the
   evolution of their relativistic jets with cosmic epoch.}
   {A complete sample of compact extragalactic radio sources at
   15\,GHz was recently compiled to conduct the MOJAVE program. The
   MOJAVE sample comprises 133 radio-loud flat-spectrum AGN with
   compact relativistic outflows detected at parsec scales.}
   {Analysis of the population of flat-spectrum quasars of the sample
   reveals that the pc-scale jets of quasars have intrinsic
   luminosities in the range between $\sim10^{24}\,{\rm W\,Hz^{-1}}$
   and $\sim10^{27}\,{\rm W\,Hz^{-1}}$ and Lorentz factors distributed
   between $3\la\gamma \la30$. We find that the apparent speed (or
   Lorentz factor) of jets evolves with redshift, increasing from
   $z\sim0$ to $z\sim1$ and then falling at higher redshifts
   ($z\sim2.5$) by a factor of 2.5. The evolution of apparent speeds
   does not affect significantly the evolution of the beamed
   luminosity function of quasars, which is most likely to be
   dependent on the evolution of radio luminosity. Furthermore, the
   beamed radio luminosity function suggests that the intrinsic
   luminosity function of quasars has a double power-law form: it is
   flat at low luminosities and steep at high luminosities. There is a
   positive evolution of quasars at low redshifts ($z<0.5$) and strong
   negative evolution at redshifts $>1.7$ with space density decline
   up to $z\sim2.5$. This implies that the powerful jets were more
   populous at redshifts between $0.5$ and $1.7$. We show that the
   evolution of compact quasars is luminosity dependent and it depends
   strongly on the speed of the jet suggesting that there are two
   distinct populations of quasars with slow and fast jets which
   evolve differently with redshift.}
   {}

   \keywords{surveys -- quasars: general -- galaxies: active -- galaxies: 
             nuclei -- galaxies: jets -- BL Lacertae objects: general
            }

   \titlerunning {Cosmological evolution of compact AGN}
   \authorrunning {T. G. Arshakian et al.}
   \maketitle


\section{Introduction}
The advantage of studying the cosmological evolution of radio
sources free of dust obscuration was realized by \cite{longair66}. He
demonstrated that the cosmic evolution of low and high luminosity
radio sources sampled at low radio frequency (178\,MHz) are different
with later ones having much stronger evolution. The morphological
difference between FRI and FRII radio sources \citep{fanaroff74} and
their division into low and high luminosity classes below and above
$L_{\rm 178\,MHz}\approx10^{25}$\,W\,Hz$^{-1}$\,sr$^{-1}$ was used by
\cite{wall80} to show that the population of FRI radio sources does
not evolve strongly with cosmic time whilst the evolution of FRII
sources is much stronger.  The orientation-based unification scheme
\citep{scheuer87,barthel89} suggests that a single parent population of radio
sources appears as radio galaxies and quasars when viewed at different
angles to the line of sight, with quasars beeing seen from
$45^{\circ}$ to the line of sight and radio galaxies from larger
angles. It was successfully used to describe a diverse appearance of
observed characteristics of radio sources and their evolution
\citep{urry95,jackson99,grimes04} at different radio frequencies. The
orientation of the jet and its beamed emission are two important keys
to identify that the BL Lacs and quasars are beamed parent objects of
FRI and FRII radio sources.  At low frequencies (178\,MHz), these
sources appear as extended (lobe-dominated) steep-spectrum FRI and
FRII radio sources. \cite{willott01} used the combined sample of 3CRR,
7C and 6CE surveys (see references in Willott et al. 2001) to model
the radio luminosity function of FRI/FRII radio sources. They showed
that a dual-population model of luminosity function fits the data
well, but it requires a differential positive density evolution
between $z\sim0$ and $z\sim2$. No evidence of a negative evolution
(redshift `cut-off') is found at high redshifts.

The surveys at frequencies $\ge1.4$\,GHz include both lobe- and
core-dominated sources in which the jet beaming effect appears to be
dominant. These radio sources enhanced by Doppler beaming of the jet
are identified with flat-spectrum quasars and BL Lacs. At these
frequencies, the cosmological evolution of flat-spectrum sources is
more complex, involving the evolution of both non-beamed and beamed
emission. In addition to the luminosity/density evolution it also may
depend on the evolution of relativistic effects with cosmic time. The
evolution of flat-spectrum quasars was studied in several works
\citep{dunlop90,jackson99,cirasuolo05,wall05}. \cite{jackson99} showed
that the evolution and beaming of powerful radio sources may be
described by a dual-population unification scheme for FRI and FRII
sources which, depending on the viewing angle, appear as
lobe-dominated steep-spectrum and core-dominated flat-spectrum
sources. Using data from a combined sample of radio sources at
2.7\,GHz \cite{dunlop90} showed that both pure luminosity and
luminosity/density evolution fit the observed redshift and
source-count data. They presented clear evidence of a redshift cut-off
for both steep- and flat-spectrum quasars at high redshifts,
$z\ga2$. A high redshift cut-off is also evident in the Parkes
quarter-Jansky flat-spectrum quasars
\citep{wall05} and in the sample of 352 faint flat-/steep-spectrum
quasars \citep{cirasuolo05} -- with radio fluxes $\ge$1~mJy at
1.4\,GHz -- selected from the radio (Faint Images of the Radio Sky at
Twenty centimeters) and optical (2dF QSO Redshift Survey)
surveys. They found an indication of a significant negative evolution
for faint quasars at $z>1.8$ consistent with negative evolution of
bright flat-spectrum radio quasars at 2.7\,GHz, and inconsistent with
no-evolution of steep-spectrum bright radio quasars at 151\,MHz
\citep{willott98}. The reduction of the comoving space density of
quasars at high redshifts is also found in X-ray and optically
selected samples (e.g., Schmidt et al. 1991; Fan et al. 2001 and
references therein; Hasinger 2004; Silverman et al. 2005) indicating
that the redshift cut-off of optically-selected quasars is not a
selection effect due to obscuration by dust.

It is not clear whether the redshift cut-off of flat-spectrum sources
is present in high radio frequency samples selected on the basis of
beamed emission and whether the cosmological evolution of beamed
sources depends on the properties of the jets such as the intrinsic
luminosity and Lorentz factor of the jet (the analytical relationships
between the latter parameters were studied by \cite{lister97} for 
simulated flux-limited samples of relativistic jets).
To examine these problems we use the (first) complete sample of 133
bright AGN at 15\,GHz which has been compiled recently by
\cite{lister05} as an extension of the Very Long Baseline Array (VLBA)
2\,cm survey \citep{kellermann98,zensus02} and other programs.  This
sample of flat-spectrum AGN was designed for a long-term observational
program (named the MOJAVE survey) to study the structure and evolution
of compact jets. The radio emission of AGN originates in the compact
jets at scales from a few parsecs to tens of parsecs. Although the
sample is small, includes relatively bright sources and is restricted
to $z\la2.7$, it provides a wealth of information about radio
emission, structure and kinematics of parsec-scale jets. The
completeness of the sample makes it possible to investigate the
cosmological evolution of flat-spectrum AGN in the context of
evolution of the jet parameters. In Sect. 2, the complete sample of
flat-spectrum AGN is presented. In Sect. 3, we investigate the
homogeneity of the MOJAVE sources on the sky. The source counts of
AGN, evolution of apparent speeds of jets, and luminosity function and
its evolution with cosmic epoch are examined in Sect. 4. Section 5
involves the summary of results. In this paper we use a flat cosmology
($\Omega_{\rm m} +
\Omega_{\rm \Lambda}= 1$) with non-zero lambda, $\Omega_{\rm m} = 0.3$
and $\rm H_{\rm 0} = 70\, km\,s^{-1}\,Mpc^{-1}$. 




\section{The MOJAVE sample}
\begin{figure}
   \centering
   \includegraphics[angle=-90,width=8cm]{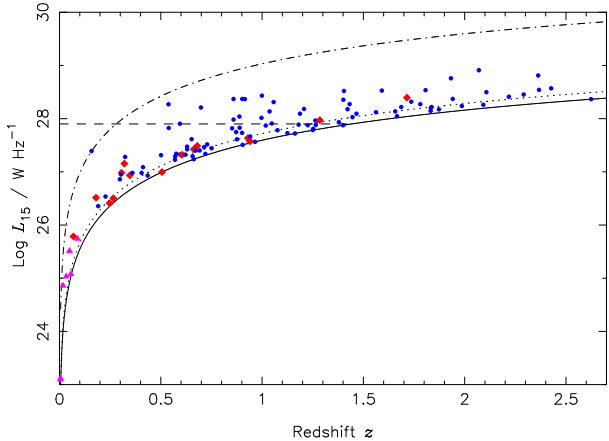}
      \caption{The apparent luminosity at 15\,GHz versus redshift for
      the 116 AGN with known redshifts. The filled circles, diamonds
      and triangles are quasars, BL Lacs and galaxies,
      respectively. The solid and dotted lines show the flux density
      limits for northern (1.5\,Jy) and southern (2\,Jy) radio sources
      respectively, and the dotted-dashed line corresponds to the
      maximum flux density (41\,Jy at a redshift 0.158) observed in
      the MOJAVE sample.}
         \label{fig1}
   \end{figure}

The main selection criterion is the flux density limit at 15\,GHz: all
variable sources with measured VLBA flux densities exceeding 1.5\,Jy
(2\,Jy for southern sources) at any epoch since 1994 are included in
the sample \citep{lister05}. It is restricted to declinations
larger than $-20°^{\circ}$ and to Galactic latitudes $|b| >
2.5^{\circ}$.  There are 96 sources in common with the 2\,cm survey, 
which is not entirely complete down to the flux density limit. 
A sample of additional 37 AGN was assembled from other recent
high-frequency surveys. The final sample comprises 133 radio sources,
all active galactic nuclei: radio-loud, core-dominated and
flat-spectrum\footnote{ http://www.physics.purdue.edu/astro/MOJAVE}
(see Lister \& Homan 2005, for more detailed description of the
sample and its completeness). The MOJAVE sample is based on the
selection of highly beamed radio sources. The majority of the AGN have
a \emph{core-jet} structure on parsec-scales, and many of them show
superluminal motion in their jets suggesting that the bulk of the
beamed emission at 15\,GHz originates in the relativistic jet. Some
nearby radio galaxies with two sided jets are included in the sample
because of bright (unboosted) radio emission. At present, 116
(87\,\%) of the 133 AGN have spectroscopic redshifts: 94 of 94
(100\,\%) quasars, 15 of 22 (68\,\%) BL Lacs and 7 of 7 (100\,\%)
radio galaxies. The remaining nine sources have no optical
identification. The radio luminosities of compact AGN at 15 GHz
($L_{15}$) vary over $(10^{23}$ to $10^{29})$ W\,Hz$^{-1}$ over the
redshift range from $0$ to $2.7$ (Fig.~\ref{fig1}). The apparent
speed of the fastest component of the jet rated as `good' or
`excellent' \citep{kellermann04} is measured for 66 quasars, 15 BL
Lacs and six radio galaxies. The maximum total flux densities observed
for each source during the VLBA monitoring campaign are used
throughout the paper.

%

\section{Sky distribution}

\begin{figure}
   \centering
   \includegraphics[angle=-90,width=8.7cm]{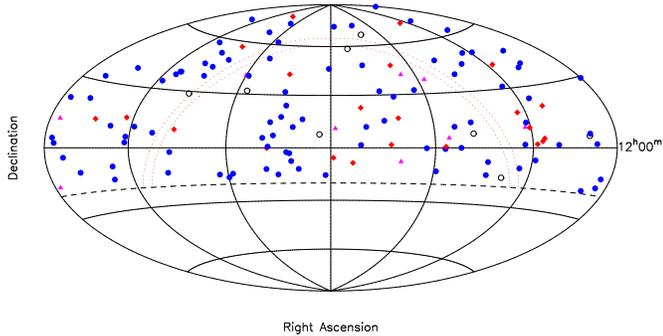}
      \caption{The distribution of 133 AGN in the region of
	       the sky restricted by $\delta > -20^{\circ}$ (the
	       dashed line). The open circles are sources with no
	       optical counterpart, other symbols are the same as in
	       Fig.~\ref{fig1}. The dotted streams indicate the
	       galactic plane exclusion strip, $|b| >
	       2.5^{\circ}$. }
         \label{fig2}
   \end{figure}

\begin{figure}
   \centering
   \includegraphics[angle=0,width=8cm]{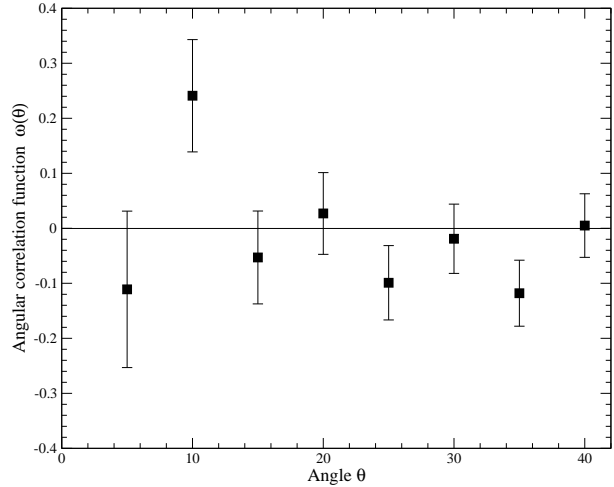}
      \caption{The correlation function against angular
      separation for 133 AGN. The uncertainty in $\omega(\theta)$
      in each bin is given by the Poisson error,
      $\sqrt{1+\omega(\theta)}/\sqrt{N_{\rm gg}}$ (Infante 1994).}
         \label{fig3}
   \end{figure}

If the sample is free from selection effects, then we should expect
that the radio sources are distributed uniformly in the sky.  The
distribution of 133 AGN on the celestial sphere is shown in
Fig.~\ref{fig2}.  The distribution of radio sources looks patchy with
few clusters and voids in the sky. 
To test the uniform distribution of AGN we measure a two-point angular  
correlation function \citep{peebles80,infante94}, 
\begin{equation}
  \omega(\theta)=\frac{2N_{\rm gg}N_{\rm r}}{N_{\rm gr}(N_{\rm g}-1)}, 
\end{equation}
where $N_{\rm g}$ is the number of AGN in the MOJAVE survey, $N_{\rm
gg}$ is the number of distinct pairs of AGN with angular separations
from $\theta$ to $\theta+\Delta\theta$, $N_{\rm r}$ is the number of
random points generated in the area of the survey, and $N_{\rm gr}$ is
the total number of random points in the same annuli $\theta$ to
$\theta+\Delta\theta$ around AGN. Note, that only $4$\,\% of the
sources ($2\pi d/4\pi$, where $d=5^{\circ}$) are missed from the
sample as a result of the Galactic plane exclusion. This has little
effect on number statistics of AGN therefore we do not take it into
account. The angular correlation function for 133 AGN is shown in
Fig.~\ref{fig3}. The correlation function remains near
$\omega(\theta)=0$ within the limits of errors, which is indicative
that the distribution of MOJAVE sources is uniform over the large
angular scale from $5^{\circ}$ to $40^{\circ}$.

%
%

\section{Space distribution: evolution of compact AGN}
At high radio frequencies the observed flux density is produced
predominantly in the relativistic jet,
\begin{equation}
\label{eq:flux}
    S=\frac{L}{4\pi R^2}(1+z)^{\alpha-1},
\end{equation}
where $L$ is the beamed (apparent) luminosity, $R$ is the comoving
distance at the redshift $z$, and $\alpha$ is the spectral index
($\alpha=0$ is assumed throughout the paper). The relativistic Doppler
effect causes the rest-frame (intrinsic) luminosity (${\cal L}$) of
the source to appear enhanced towards the observer, $L=\delta^p{\cal
L}$, where $p=2-\alpha$ for a steady-state jet and $\delta$ is the
Doppler factor which is a function of the jet viewing angle ($\theta$)
and the speed of the jet $\beta=(1-1/\gamma^2)^{1/2}$ (in units of the
speed of the light) and $\gamma$ is the Lorentz factor,
\begin{equation}
\label{eq:delta}
    \delta=\frac{1}{\gamma(1-\beta\cos\theta)}.
\end{equation}
Therefore, the source counts and cosmological evolution of AGN
from the MOJAVE sample are affected by the relativistic beaming
effects. 

\subsection{Source counts}
The MOJAVE sample is sampled from different high frequency radio
samples and does not represent an independent survey. One needs to
test its completeness which is the main criterion for deriving the
true cosmological evolution and statistical characteristics of
relativistic jets of compact AGN. If the MOJAVE sample is a complete
sample, then one should expect that the slopes of source counts of the
MOJAVE sample and independent high frequency surveys are aligned. At
present, the only independent survey of radio sources at 15.2\,GHz is
carried out using the Ryle Telescope
\citep{taylor01}. They identified 66 sources between 0.2\,Jy and 5\,Jy
and fitted the differential source count by a power law, $n(S)\propto
S^{-2}$. At 15.2\,GHz and flux density range the main population of
objects should consist of the beamed flat-spectrum extragalactic radio
sources \citep{jackson99}.

\begin{figure} 
     \centering \includegraphics[width=8.5cm]{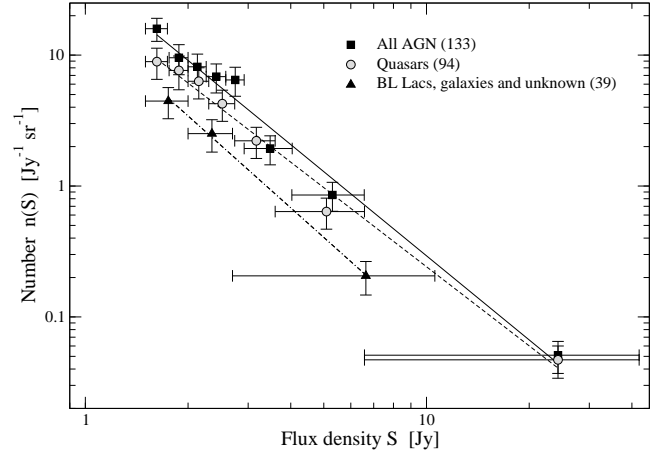}
     \caption{The differential source count of AGN sampled at 15
              GHz. Filled squared represent all radio sources, grey
              circles are quasars and filled diamonds are BL Lacs,
              galaxies and unknown type of sources. The length of
              horizontal bars marks the flux density bin size. }
     \label{number-counts} 
   \end{figure}

The normalized differential source count of AGN from the MOJAVE sample
is presented in Fig.~\ref{number-counts}. To take into account the
different flux density limits of the sample, the northern $2\pi$ sky
area is used to normalize the number count of AGN between the flux
density range from 1.5\,Jy to 2\,Jy, and the number count of AGN with
$S>2$\,Jy is normalized to the complete sky area of the sample,
$2\pi(1+0.342)$. The slope of the differential source counts
($n(S)\propto S^{a}$\,Jy$^{-1}$sr$^{-1}$) is estimated for three sets
of sources (Fig.~\ref{number-counts}): all AGN ($a=-2.15\pm0.11$),
quasars ($a=-2.01\pm0.09$), and BL Lacs, galaxies and sources with no
optical identification ($a=-2.33\pm0.08$). The MOJAVE and 15.2\,GHz
surveys overlap between the 1.5\,Jy and 5\,Jy flux density range. The
source counts of all MOJAVE sources (solid line) and quasars only
(dashed line) have slopes which are characteristic of the complete
survey at 15.2\,GHz ($a=-2$) and of the bright end ($>1$\,Jy) of
lower frequency surveys at 0.15\,GHz to 8.4\,GHz ($a\sim-1.9$)
suggesting that the MOJAVE represents a flux-limited complete
sample. The subsample of non-quasars (dot-dashed line) has a slightly
steeper slope. It is most likely that this deviation arises because of
a small number of bins (or sources) involved in the fitting.

It is useful to compare the observed source count with the Euclidean
source count which is given by $N(>S)\propto S^{-3/2}$ for the
unbeamed and uniformly distributed radio sources. It is not obvious
whether this simple relation holds for the beamed sample of radio
sources. Given the apparent radio luminosity function $\Phi(L,z)$, the
number of radio sources with apparent flux $>S$ per unit area of the
sky is given by,
\begin{equation}
\label{eq:ncount1}
  N(>S)=\frac{1}{4\pi}\int_{L_{\rm min}}^{L_{\rm
  max}}{dL}\int_{0}^{z(S,L)}{\Phi(L,z)\frac{dV(z)}{dz}dz},
\end{equation}
where $dV$ is the comoving volume element. For the isotropic
non-evolving Euclidean universe $dV=4\pi R^2dR$ and $S=L/(4\pi R^2)$,
and the equation~(\ref{eq:ncount1}) reduces to,
\begin{equation}
\label{eq:ncount2}
   N(>S)\propto S^{-3/2}\int_{L_{\rm min}}^{L_{\rm max}}{L^{3/2}\Phi(L)dL}.
\end{equation}
The apparent luminosity function is the convolution of the intrinsic
luminosity function ($\psi({\cal L})$) and probability density of a 
Doppler factor ($P_{\rm \delta}(\delta)$),
\begin{equation}
\label{eq:RLF}
   \Phi(L)dL=\frac{L^{1/p-1}}{p}dL\int\psi({\cal L})P_{\rm
   \delta}(L/{\cal L})d{\cal L},
\end{equation}
where the integration limits are a complex function of $L$ and
$\delta$ \citep{urry91,lister03}.  The integral over the apparent
luminosity function is constant (eq.~\ref{eq:ncount2}) and hence the
Euclidean differential source counts of a beamed population of radio
sources, $n(S)=dN(>S)/dS\propto S^{-2.5}$, is independent of the
distributions of their intrinsic luminosities and Doppler factors. The
slope of the differential number counts of the MOJAVE sources
($a=-2.1$) is flatter than the slope of the Euclidean source counts
($a=-2.5$) indicating that the radio luminosity function of compact
AGN evolves with cosmic epoch.

\subsection{Radio luminosity function}
The observed radio luminosity of BL Lacs and quasars at 15\,GHz is
highly boosted and radiated predominantly by the approaching
relativistic jet \citep{kellermann98,zensus02}. The relativistic
Doppler beaming of the jet transforms the intrinsic radio luminosity
function of compact quasars to the beamed (apparent) luminosity
function at the rest-frame of the observer. Earlier studies
\citep{urry84,urry91,lister03} provided analytical predictions for the
apparent luminosity function of jets randomly oriented on the sky.
The single and double power-law intrinsic luminosity functions were
modelled assuming a single-valued Lorentz factor for the
jet. \cite{lister03} extended the later work for the variety of
power-law distributions of Lorentz factors, $P_{\gamma}\propto
\gamma^a$ ($\gamma_1\le\gamma\le\gamma_2$), which was found to provide
a good fit to the distribution of apparent speeds of AGN
\citep{lister97}.  The modelled apparent and intrinsic luminosity
functions have several important break points and relations. We will
use these to constrain the intrinsic luminosity function of the
jets. The evolution of the apparent luminosity function involves both
evolution of the intrinsic luminosity function and evolution of the
Lorentz factor (or the change of the Doppler factor) of the jet with
redshift. The latter can be examined by considering the apparent
speed--redshift relation plane which also allows the important
characteristics of the relativistic beaming of flat-spectrum AGN to be
determined.



%
\begin{figure}
   \centering
   \includegraphics[width=8.2cm]{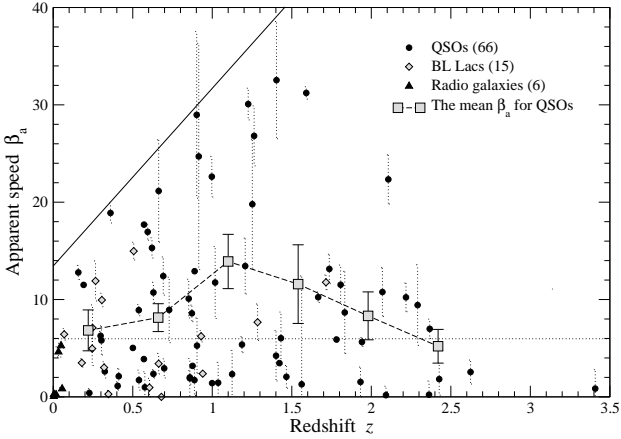} 
   \caption{Apparent transverse speed of the jet as a function of
   redshift for AGN with known values of $\beta_{\rm a}$: 66 quasars
   (filled circles), 15 BL Lacs (gray diamonds) and 6 radio galaxies
   (filled triangles). The gray squares are the mean apparent speed of
   quasars estimated in six redshift bins (binning ranges are the same
   as in Fig.~\ref{fig:zdist}). The dotted line shows the division of
   quasars in two equal subsamples with low ($\beta_{\rm a}<6$) and
   high apparent speeds ($\beta_{\rm a}\ge6$). 1\,$\sigma$ error are
   presented for all data. }
   \label{fig:va-z}
\end{figure}
\begin{figure}
   \centering
   \includegraphics[angle=-90,width=8.2cm]{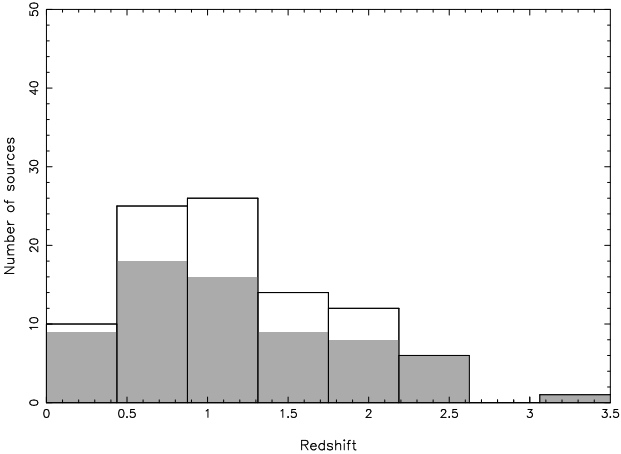}
   \caption{The cumulative redshift distribution of compact
   quasars. The $z$-distribution of 66 quasars with known values of
   apparent speed (grey area) is superimposed with the distribution of
   28 quasars with unknown $\beta_{\rm a}$ (clear area).}
   \label{fig:zdist}
\end{figure}
\subsubsection{The apparent speed -- redshift plane}
The apparent speed $\beta_{\rm a}$ (in units of the speed of the
light, $c$) of jets measured for 87 AGN is plotted against their
redshifts in Fig.~\ref{fig:va-z}. The mean apparent speed varies
significantly depending on type of AGN: ($1.9\pm 1.0)c$ for radio
galaxies, ($5.9\pm1.1)c$ for BL Lacs and ($9.4\pm1.0)c$ for
quasars. Most of quasars and BL Lacs have superluminal jets
($\beta_{\rm a}>1$) indicating that their radio emission is highly
beamed. This puts a limit on the minimum Doppler factor, $\delta_{\rm
min}=\delta(\gamma_1,\theta_{\rm c})=1$ ($L\ga{\cal L}$), where
$\theta_{\rm c}$ is the maximum (critical) viewing angle of the jet
having a minimum Lorentz factor $\gamma_1$. Assuming that $\theta_{\rm
c}\approx 45^{\circ}$ for quasars \citep{barthel89}, we calculate the
lower limit on the Lorentz factors, $\gamma_1\sim 3$ (from
eq.~\ref{eq:delta}).

\cite{vermeulen94} showed that for the orientation-biased sample with
a single Lorentz factor distribution the value of the apparent speed
is close to the maximum apparent speed of the jet $\beta_{\rm
a}\approx\beta_{\rm a,\,max}=\sqrt{\gamma^2-1}\approx \gamma$ which is
true for BL Lac objects and quasars with $\gamma^2\gg1$, and
$\delta_{\rm max}\approx 2\gamma_2=2\beta_{\rm a,\,max}$. The maximum
apparent speed and Lorentz factor of the jet is
$\gamma_2\approx\beta_{\rm a,\,max}\approx 30$ (Fig.~\ref{fig:va-z})
and the $\delta_{\rm max}\approx60$. The maxumum Doppler factor will
be reached at the viewing angle $\approx0$ degrees. It is extremely
unlikely to detect a source with $\delta_{\rm max}=60$ in a small
sample such as the MOJAVE. $\delta\approx\beta_{\rm a,\,max}$ at the
optimum angle of the jet, $\sin\theta=1/\gamma$, and the value of
$\delta\approx30$ is a realistic upper limit to the Doppler factor of
the jets of our sample. To calculate the range of the intrinsic
luminosities of jets we use the minimum and maximum radio luminosities
of quasars ${L}_{\rm min}=2.3\times10^{26}$\,W\,Hz$^{-1}$ (at
$z\sim0.2$) and ${L}_{\rm max}=10^{29}$\,W\,Hz$^{-1}$ (at
$z\sim2$). Assuming that $\delta\approx\beta_{\rm a,\,max}\sim10$ at
low and high redshifts (see Fig.~\ref{fig:va-z}) we recover
${\cal L}_{1}=\delta^{-2}L_{\rm min}\approx2.3\times10^{24}\,{\rm
W\,Hz^{-1}}$ and ${\cal L}_{2}=\delta^{-2}L_{\rm
max}\approx10^{27}\,{\rm W\,Hz^{-1}}$.

There is a lack of high $\beta_{\rm a}$ at low redshifts, $z<1$
(Fig.~\ref{fig:va-z}). For the $\beta_{\rm a}-L$ relation plane, Cohen
et al. (in preparation) discuss the selection effects which might
result in the lack of sources with large $\beta_{\rm a}$ at low
luminosities. They argue that both effects, the minimum sampling
interval of the MOJAVE sample and the determination of speed of the
jet when the frequency of ejection of moving components is greater
than the frequency of observations, can not account for the lack of
sources thus leading to the conclusion that the increase of apparent
speeds with increasing luminosity is real. This is true also for the
$\beta_{\rm a}-z$ relation plane because of positive correlation
between $L$ and $z$ in the flux-limited MOJAVE sample. Another
potential selection effect at low redshifts is that we miss sources
with unknown $\beta_{\rm a}$ which may populate the left-top region in
Fig.~\ref{fig:va-z}. The quasars with unknown apparent speeds
(Fig.~\ref{fig:zdist}) will populate the redshifts $z>0.5$ thus the
increase of $\beta_{\rm a,\,max}$ to high redshifts ($z\sim1$) will
not be affected. This trend is also evident for the mean apparent
speeds (gray squares in Fig.~\ref{fig:va-z}) estimated in different
redshift bins. There is a gradual decrease of the mean apparent speed
from $z\sim1$ down to $z\approx 2.5$. Were high speed jets
($\beta_{\rm a}>30$) to exist at high redshifts one should observe
them because their radio emission would be highly boosted. It is
unlikely that the high-speed quasars are among those with unknown
$\beta_{\rm a}$, because no systematic selection of measured apparent
speeds was introduced. For the same reason, the averaged behaviour of
the apparent speed can not be affected by unknown apparent speeds.
Moreover, the highest redshift bin, which includes six quasars with
measured $\beta_{\rm a}$, has the lowest value of $\langle\beta_{\rm
a}\rangle=5.2\pm1.7$ which supports the idea that the apparent speeds
(and hence the Lorentz factors) of the jets of quasars evolve with
cosmic epoch: the mean apparent speed increases from $z\sim2.5$ to
$z\sim1$ by a factor 2.5 and then falls by factor of two at the
present epoch.

Note that a peak in the $\langle\beta_{\rm a}\rangle-z$ relation plane
is mainly due to nine high speed sources with $\beta_{\rm a}>20$, and
that inclusion or exclusion of these quasars has no significant effect
on the evolution of the luminosity function of quasars examined in the
next sections.

\subsubsection{Evolution of the luminosity function of quasars}
\begin{figure}
   \centering
   \includegraphics[angle=-90,width=8.2cm]{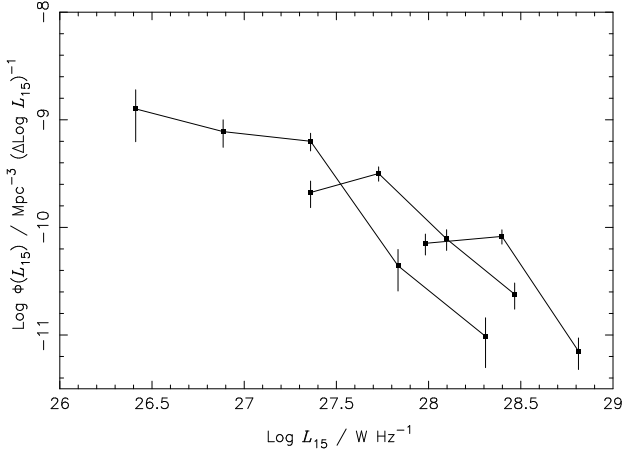}
   \caption{The radio luminosity function of 94 quasars is presented
   for three redshift bins $z<0.7$ (left curve), $0.7\le z<1.4$
   (middle curve) and $z\ge1.4$ (right curve). The Poisson error (1
   $\sigma$) acossiated with the luminosity function is given
   ($\sigma^2=(\Delta Log(L_{15}))^{-1}\sum 1/V_{\rm a}^2$).}
         \label{fig:RLF2}
   \end{figure}

We chose the population of quasars for a study of the luminosity
function because they comprise a large homogeneous sample of beamed
and optically identified sources having apparent speeds measured for
$\sim$70\% of the jets. We used the $1/V_{\rm a}$ \citep{schmidt69}
method to construct the apparent luminosity function. The MOJAVE
sample occupies different sky areas depending upon the flux density
limits: 1.5\,Jy for northern hemisphere and 2\,Jy for southern
hemisphere. In this case, the $V_{\rm a}$ is the volume
\emph{available} in the combined sample \citep{avni80}. Note that all
statistical tests and methods throughout the paper were applied to the
combined sample. The apparent luminosity function is presented in
Fig.~\ref{fig:RLF2} for three redshift bins, $0<z<0.7$ (27 quasars),
$0.7\le z<1.4$ (35) and $z\ge 1.4$ (32). The form of the luminosity
function is similar at all redshifts: it is flat at lower luminosities
($A\sim0$, for $\Phi(L)=L^{-A}$) and becomes steeper $A\sim
(1.9\,\,{\rm to}\,\, 2.5)$ at higher luminosities. There is marginal
evidence that the high luminosity end of the luminosity function
steepens with increasing redshift.
We now discuss the form of the intrinsic luminosity function and
distribution of Lorentz factors which may reproduce the observed
luminosity function. We consider a single (steep) and double (flat and
steep) power-law intrinsic luminosity functions ($\phi({\cal L})={\cal
L}^{-B}$, where ${\cal L}_1\le{\cal L}\le{\cal L}_2$) and distribution
of Lorentz factors with $\gamma_1\le\gamma\le\gamma_2$
\citep{urry91,lister03}.


%
\cite{urry91} showed that if the slope of the apparent luminosity
function is steeper than $-(p+1)/p=-1.5$ ($p=2$) then the intrinsic
and apparent luminosity functions have the same slope at high
luminosities $L>L_4=\delta_{\rm max}^p{\cal L}_1$ (here we use the
notation of Urry \& Padovani 1991). At high luminosities the slope of
the apparent luminosity function (and hence the slope of the intrinsic
luminosity function) is $\sim2.5$. \cite{urry91} claimed that such a
single and steep power-law intrinsic luminosity function will be
transformed to the apparent luminosity function which, at low
luminosities ($L<L_4$), will have a slope $-(p+1)/p=-1.5$ independent
of the Lorentz factor distribution. A single power-law parent
luminosity function can not produce the flat part ($B\sim0$) of the
apparent luminosity function observed at low
luminosities. \cite{lister03} showed that a flat or inverted slopes
may be produced with the steep power-law slope ($B=2$) of the
intrinsic luminosity function, $1\le\gamma\le 30$ and steep ($a<0$)
distribution of the Lorentz factors. This is true for
orientation-unbiased samples populated by slow jets and apparently it
is not applicable to the flux-limited MOJAVE sample which is dominated
by relativistic jets and high $\delta>1$.

A complex (flat and steep) power-law intrinsic luminosity function and
single power-law distribution of Lorentz factors of the jets are
needed to describe the observed flatness of the luminosity function
of beamed quasars. At low luminosities, the flat slopes of the
intrinsic and apparent luminosity functions should coincide
\citep{urry91} while the steep slopes should match at high
luminosities. The overall intrinsic luminosity function can be crudely
represented by a double power-law with $\phi({\cal L})\sim const$ and
$\phi({\cal L})\sim{\cal L}^{-2.5}$ at low and high luminosities
respectively. The broken power-law form is generally consistent with
the luminosity functions of quasars selected from unbeamed surveys:
optical \citep{boyle88} and low radio frequency samples
\citep{willott98,willott01}. This supports the idea that
flat-spectrum quasars are the parent population of FRII radio sources
beamed towards the observer \citep{jackson99}.


The shift of the apparent luminosity function (Fig.~\ref{fig:RLF2})
may happen as a result of the evolution of intrinsic luminosity
function of quasars and/or evolution of the Lorentz factor (Doppler
factor) of the jet with cosmic time.  We now consider three cases: a
pure evolution of the Lorentz factors of the jets, a pure evolution of
the intrinsic luminosity function, and evolution of both. The mean
Lorentz factors in three redshift bins (Fig.~\ref{fig:RLF2}) are
($8.3\pm1.4$) at low-$z$, ($11.6\pm2.2$) at intermediate-$z$ and
($8.4\pm1.8$) at high-$z$. In the case of an evolution of the jet
speeds, the unchanged intrinsic luminosity function will be
transformed to the apparent luminosity function. The observed shift of
the luminosity function at high redshifts requires the increase of the
beaming factor (or the Lorentz factor) with increasing redshift. The
mean Lorentz factor increases from low-$z$ ($8.3\pm1.4$) to
intermediate-$z$ ($11.6\pm2.2$) and goes down at high-$z$
($8.4\pm1.8$). This is in contradiction with the expected increase of 
the beaming factor at high redshifts.   

In the case of a pure luminosity evolution of the intrinsic luminosity
function one should expect that the luminosity curves shift to high
luminosities (Fig.~\ref{fig:RLF2}) at high redshifts (similar to the
shift of the luminosity function of optically selected quasars,
e.g. Boyle et al. 1988). \cite{wall05} noted that a contamination of
the low luminosity bin by unbeamed sources can mimic a luminosity
evolution. Our sample includes only the beamed quasars and therefore
the observed shift is likely to be produced by simple luminosity
evolution. An important question is whether the evolution of the
beaming effects have a contribution to the shift of the luminosity
function. Although there are indications that the Lorentz factor
evolves with cosmic time (Fig.~\ref{fig:va-z}) it has no significant
effect on the evolution of the apparent luminosity function. To
understand and recover fully the effects of the relativistic beaming
on the intrinsic luminosity function one needs to compare the observed
distributions of jet parameters (and a complete data set of apparent
speeds) with the distributions generated by means of Monte-Carlo
simulations (Lister \& Marscher 1997; Lister et al., in preparation).

\subsection{$V/V_{\rm max}$ evolution}
Here, we use the model-independent $V/V_{\rm max}$ statistics
\citep{schmidt69} to test a null hypothesis of a uniform distribution
of compact AGN in space and to study their cosmological
evolution. This test is applicable only for complete flux-limited
samples, where $V$ is the volume of space enclosed by the redshift of
the AGN having a certain flux density, and $V_{\rm max}$ is the
maximum volume of space within which this source could be observed and
still be included in the complete sample. If sources have a uniform
distribution in space, then the averaged ratio of $V/V_{\rm max}$ is
constant and equal to 0.5.  The difference between the expected and
observed values of $V/V_{\rm max}$ allows the significance of a
deviation of the real distribution from uniform to be estimated.  This
test has been applied for different radio samples to show the presence
of a strong positive evolution up to $z\sim2$ with $\langle V/V_{\rm
max} \rangle \sim 0.6$ to $0.7$
(e.g. \citep{longair70,wills78,peacock81}.
  
We employed the generalized $V/V_{\rm max}$ test \citep{avni80} which
is derived for combined samples with different flux-density limits and
sky areas. The new variables are $V_{\rm e}$ and $V_{\rm a}$, the
volume \emph{enclosed} by the source and the volume \emph{available}
in the combined sample.  \cite{avni80} showed that the formal
statistical properties of $V_{\rm e}/V_{\rm a}$ are the same as those
of $V/V_{\rm max}$.

For 94 quasars, $\langle V_{\rm e}/V_{\rm a} \rangle = 0.589 \pm
0.027$, the distribution is not uniform at the $99.3$\,\% confidence
level (see Table 2), and it is biased towards large values which
indicates that compact quasars, and hence the powerful jets, were more
populous at high redshifts.  \begin{table}
\caption[]{$V_{\rm e}/V_{\rm a}$ test for 116 AGN.}
\label{vvmax_test} \center{ \begin{tabular}{lcclll} \hline
\noalign{\smallskip} ID & Num & $\langle V_{\rm e}/V_{\rm a} \rangle$
& Standard & \multicolumn{2}{c} {K-S test} \\ & $N$ & & Error $^{\rm
a}$ & (\%) & Prob. \\ \noalign{\smallskip} \hline \noalign{\smallskip}
All & 116 & 0.581 & 0.025 & 98.83 & 0.012 \\ Quasars & 94 & 0.589 &
0.027 & 99.02 & 0.009 \\ BL Lacs & 15 & 0.617 & 0.070 & 52 & 0.48 \\
Galaxies & 7 & 0.375 & 0.102 & 68 & 0.32 \\ \noalign{\smallskip}
\hline \end{tabular}
\begin{list}{}{}
\item[$^{\mathrm{a}}$] The standard error is estimated from SE $=
\sqrt{\langle x^2 \rangle - {\langle x \rangle}^2 }/\sqrt{N}$, where
$x=V_{\rm e}/V_{\rm a}$ and $N$ is the number of sources.
\end{list}
} \end{table} The BL Lacs show a similar trend, but this is not
statistically significant because of the small number of sources. The
distribution seems to be uniform for galaxies, i.e. with no evolution
so far. The plausible explanation is that all eight radio galaxies
occupy the low-redshift region where the density/luminosity evolution
is negligible, but better statistics would be needed to confirm this
result.

To investigate the dependence of radio luminosity on the $V_{\rm
e}/V_{\rm a}$ statistics, we divided the sample of quasars in two
equal subsamples by absolute luminosity at 15\,GHz, $L_{15} =
10^{27.9}\,{\rm W\,Hz^{-1}}$. Strong sources have redshifts greater
than 0.5 and weak sources have $z<1.5$ (Fig.~\ref{fig1}). For 46
low-luminosity quasars we find that $\langle V_{\rm e}/V_{\rm a}
\rangle = 0.658\pm0.036$ with confidence level of $99.96$\,\%,
indicative that the distribution of $V_{\rm e}/V_{\rm a}$ is biased
towards large values, whilst for 48 strong sources, $\langle V_{\rm
e}/V_{\rm a} \rangle = 0.53\pm0.04$ (41\,\%), no significant deviation
from a uniform distribution was found. The K-S test rejected at 96\,\%
confidence level the null hypothesis that these distributions are
drawn from the same parent population. The Student t test rejected the
hypothesis of equal mean values of $V_{\rm e}/V_{\rm a}$ for low- and
high-luminosity quasars at high significance level, 0.011
(98.88\,\%). This suggests that the cosmological evolution of quasars
depends on radio luminosity.
The evolution of high-redshift quasars can be masked by the strong
positive evolution at low redshifts. To investigate the evolutionary
behaviour of quasars at high redshifts we used the banded version of
the $V/V_{\rm max}$ test \citep{osmer70,avni83} which allows one to
mask out evolutionary effects at high redshifts. It calculates the
mean of a new statistic, $(V_{\rm e}-V_{\rm 0})/(V_{\rm a}-V_{\rm
0})$,
for the samples restricted by different values of $z_{0}$ ($z > z_{0}$),
where $V_{\rm 0}$ is the volume enclosed by a redshift $z_{0}$.

\begin{figure}
   \centering
   \includegraphics[angle=-90,width=8.5cm]{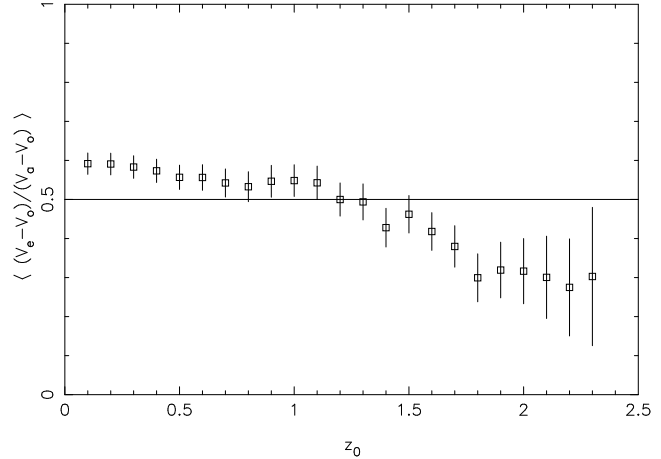}
      \caption{The banded $V/V_{\rm max}$ statistic $\langle (V_{\rm
      e}-V_{\rm 0})/(V_{\rm a}-V_{\rm 0}) \rangle$ against $z_{\rm
      0}$, the lower limit in the redshift bins. This plot is presented
      numerically in the Table 3. 1$\sigma$ standard
      error (see Table 3) is presented.}
         \label{fig:vvm-b}
   \end{figure}

\begin{figure}
   \centering
   \resizebox{\hsize}{!}{\includegraphics[angle=-90]{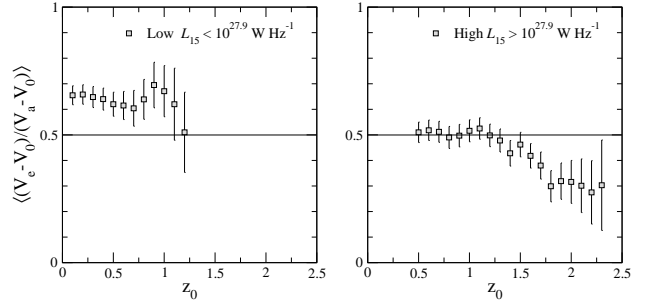}}
      \caption{The banded $V/V_{\rm max}$ statistic versus $z_{\rm 0}$
      for 46 low-luminosity (left panel) and 48 high-luminosity (right
      panel) quasars.}
         \label{fig:vvm-b-l}
   \end{figure}

\begin{figure}[b]
   \centering
   \resizebox{\hsize}{!}{\includegraphics[angle=-90]{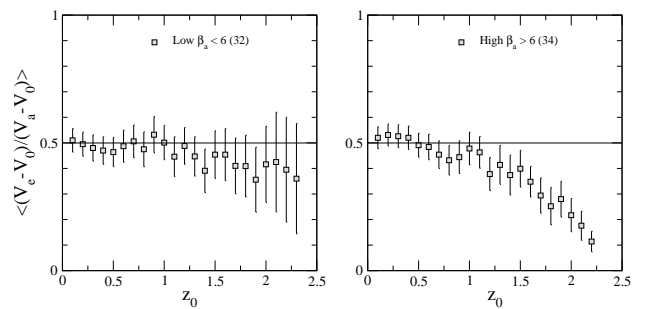}}
      \caption{The banded $V/V_{\rm max}$ statistic versus $z_{\rm 0}$
      for 32 low-$\beta_{\rm a}$ (left panel) and 34 high-$\beta_{\rm
      a}$ (right panel) quasars with known apparent speeds.}
         \label{fig:vvm-speed}
   \end{figure}

The results of a banded $V/V_{\rm max}$ test are shown in
Figs.~\ref{fig:vvm-b}-\ref{fig:vvm-speed}. For 94 quasars, a weak
positive evolution is seen at low redshifts, $z < 1 $, with values
greater than the 0.5 no-evolution line (Fig.~\ref{fig:vvm-b}). The
positive evolution is significant for quasars with $z < 0.2$ (see
Table 3). The values of $\langle V_{\rm e}/V_{\rm a} \rangle$ decrease
gradually to 0.5, and, within the errors, they remain at no-evolution
level out to $z \sim 1.6$. A negative evolution is becoming
significant for quasars with $z > 1.7$ (Table 3), and it remains
significant out to $z \sim 2.5$ at which only few sources can be
found. A significant redshift cut-off indicates that the population of
beamed quasars having powerful jets undergo a diminution already at
$z\sim1.7$. Slightly high redshift cut-off ($z>1.8$) is found for
faint quasars sampled at 1.4\,GHz
\citep{cirasuolo05} and for the combined flat-spectrum quasars at
2.7\,GHz ($z>2$, Dunlop \& Peacock 1990).

   \begin{table}
      \caption[]{Tabulated $\langle V_{\rm e}/V_{\rm a} \rangle$
      results for 94 flat-spectrum quasars.}
         \label{vvmax_banded_table} 
	 \center{
         \begin{tabular}{lccl}
            \hline
            \noalign{\smallskip}
            Redshift & Number & $\langle V_{\rm e}/V_{\rm a} \rangle \pm$ SE$^{\rm a}$ & {K-S test}  \\
	    range    & $N$    &                                                        & Prob.             \\
            \noalign{\smallskip}
            \hline
            \noalign{\smallskip}
            $>$0.1 &  94 & 0.592 $\pm$ 0.027 &  0.008 \\
            $>$0.3 &  90 & 0.583 $\pm$ 0.029 &  0.047 \\
            $>$0.5 &  84 & 0.557 $\pm$ 0.031 &  0.280 \\
            $>$0.7 &  67 & 0.542 $\pm$ 0.036 &  0.361 \\
            $>$0.9 &  56 & 0.547 $\pm$ 0.041 &  0.295 \\
            $>$1.1 &  43 & 0.543 $\pm$ 0.043 &  0.383 \\
            $>$1.3 &  33 & 0.494 $\pm$ 0.046 &  0.937 \\
            $>$1.5 &  25 & 0.462 $\pm$ 0.048 &  0.352 \\
            $>$1.7 &  20 & 0.380 $\pm$ 0.053 &  0.021 \\
            $>$1.9 &  13 & 0.319 $\pm$ 0.071 &  0.009 \\
            $>$2.1 &   8 & 0.301 $\pm$ 0.105 &  0.012 \\
            $>$2.3 &   5 & 0.303 $\pm$ 0.177 &  0.114 \\
            \noalign{\smallskip}
            \hline
         \end{tabular}
	 \begin{list}{}{}
	 \item[$^{\mathrm{a}}$] The standard error: SE $=
	   \sqrt{\langle x^2 \rangle - {\langle x \rangle}^2
	   }/\sqrt{N}$, where \\$x=(V_{\rm e}-V_{\rm 0})/(V_{\rm
	   a}-V_{\rm 0})$ and $N$ is the number of sources with $z>z_0$.
	 \end{list}
	 }
   \end{table}

Figure~\ref{fig:vvm-b-l} shows the results of a banded test for low-
and high-luminosity quasars separated by luminosity at $L_{\rm
15\,GHz} = 10^{27.9}\,{\rm W\,Hz^{-1}}$. There is significant positive
evolution for low-luminosity quasars out to $z \sim 0.5$, and, again,
a strong evidence for a high-redshift decline in the comoving space
density of high-luminosity quasars with $z \geq 1.7$. The luminosity
appears to play an important role in the evolution of flat-spectrum
quasars.

We find that the evolution of flat-spectrum quasars depends
strongly on the speed of the jet. Relatively slow jets ($\beta_{\rm
a}<6$, see Fig.~\ref{fig:va-z}) do not show significant evolution up
to $z\approx2.5$ (Fig.~\ref{fig:vvm-speed}), whilst the population of
fast jets grows rapidly between $z\sim2.5$ and $z\sim1.7$ and remains
at no evolution level at small redshifts.  As discussed in Sect.
4.2.1, the missed quasars with no $\beta_{\rm a}$ values should
populate randomly the samples of high- and low-speed quasars and
hence it is unlikely that their exclusion will introduce the observed
differential evolution of quasars with high and low speeds. For a
given apparent luminosity the population of slow jets should have
brighter intrinsic luminosities than those of slow jets. This hints
that there should be two distinct populations of quasars, one with fast
jets and low intrinsic luminosities which forms at later epochs and
undergoes a strong evolution up to $z\sim1$ and one with slow jets and
high intrinsic luminosities formed at much earlier cosmic epochs and
evolving slowly to present time. A dual population model consisting of
an \emph{old} population of slowly evolving low speed jets and
\emph{young} population of strongly evolving fast jets at $z>1$ may
produce an apparent `negative evolution' of $\beta_{\rm a}$ seen at
$z>1$ (Fig.~\ref{fig:va-z}). In this scenario, the positive evolution
of the apparent speed at $z\la1$ (Fig.~\ref{fig:va-z}) is mainly due
to young population of high speed quasars.

%

%



%

\section{Conclusions and discussions}
The principal results of statistical analysis of bright flat-spectrum
compact AGN from the flux-limited MOJAVE sample are: (a) the
distribution of 133 AGN is uniform on the sky; (b) the slope of the
source counts of radio sources is aligned with the slopes of source
counts of independent high radio frequency surveys suggesting that the
MOJAVE is a \emph{complete} flux-limited sample; (c) Lorentz
factors of the jets of quasars lie in the range from $\sim3$ to
$\sim30$ and Doppler factors of the jets are distributed between
$\delta_{\rm min}>1$ and $\delta_{\rm max}\la60$ with the realistic
upper limit on $\delta_{\rm max}\approx30$ for the MOJAVE sample; (d)
the mean apparent transverse speed of jets of quasars (or the mean
Lorentz factor) increases by a factor of two (from 6.5 to 14) with
increasing redshift up to $z\sim1$ and then gradually decreases at
high redshifts by a factor of 2.5; (e) the intrinsic luminosities of
compact jets are distributed in the range from $\sim10^{24}\,{\rm
W\,Hz^{-1}}$ to $\sim10^{27}\,{\rm W\,Hz^{-1}}$ and their intrinsic
radio luminosity function is flat ($B\sim0$) at low luminosities and
becomes steeper ($B\sim-2.5$) at high luminosities; (f) the weak
positive evolution of quasars is evident at low redshifts, $z<0.5$,
and a statistically significant redshift cut-off is present at $z \geq
1.7$; (g) the cosmological evolution is different (1) for high-
($L_{15}>10^{27.9}\,{\rm W\,Hz^{-1}}$) and low-luminosity quasars
($L_{15}<10^{27.9}\,{\rm W\,Hz^{-1}}$) suggesting that their evolution
depends strongly on luminosity, and (2) for fast and slow jets, which
indicates the existence of two distinct populations of quasars one
with slow jets ($\beta_{\mathrm a}<6$) and high intrinsic luminosities
which does not show a significant evolution with redshift and one with
$\beta_{\mathrm a}>6$ and low intrinsic luminosities having a
pronounced redshift cut-off at $z>1.7$.

The differential evolution of the population of fast and slow jets at
$z>1.7$ may explain inconsistent results obtained for a redshift
cut-off of quasars sampled at high and low radio frequencies
\citep{dunlop90,willott01} with former one providing no evidence for
any decline in the comoving space density. If slow jets reside in high
luminosity FRII radio sources and fast jets produce faint FRII
structures, which are too weak to be detected at 151\,MHz, then the
low-frequency sample will include the population of quasars with low
jet speeds which do not show significant evolution up to $z\sim2.5$
(Fig.~\ref{fig:vvm-speed}). The population of fast jets, however, will
be included in the high frequency samples because of highly beamed
emission of the jet. This naturally explains the redshift cut-off of
quasars sampled at 1.4\,GHz, 2.7\,GHz
\citep{dunlop90,cirasuolo05,wall05} and significant redshift cut-off
at $z>1.7$ found for our sample
(Figs.~\ref{fig:vvm-b-l},\ref{fig:vvm-speed}).

\begin{acknowledgements}
      The authors thank Christian Wolf and members of the MOJAVE team,
      Marshall Cohen, Dan Homan, Ken Kellermann, Yuri Kovalev and Matt
      Lister for comments and useful discussions, and the referee for
      constructive suggestions which significantly improved the
      paper. TGA is grateful to the Alexander von Humboldt Foundation
      for the award of a Humboldt Post-Doctoral Fellowship.
\end{acknowledgements}

\bibliographystyle{aa} 

\begin{thebibliography}{}
\bibitem[Avni \& Bahcall(1980)]{avni80} Avni, Y., \& Bahcall,
  J. N. 1980, \apj, 235, 694
\bibitem[Avni \& Shiller(1983)]{avni83} Avni, Y., \& Schiller,
  N. 1983, \apj, 267, 1
\bibitem[Barthel(1989)]{barthel89} Barthel, P.~D.\ 1989, \apj, 336,
  606
\bibitem[Boyle et al.(1988)]{boyle88} Boyle, B.~J., Shanks, T., \&
  Peterson, B.~A.\ 1988, \mnras, 235, 935
\bibitem[Cirasuolo et al.(2005)]{cirasuolo05} Cirasuolo, M.,
  Magliocchetti, M., \& Celotti, A.\ 2005, \mnras, 357, 1267
\bibitem[Dunlop \& Peacock(1990)]{dunlop90} Dunlop, J. S., \& Peacock,
  J. A. 1990, MNRAS, 247, 19 
\bibitem[Fanaroff \& Riley(1974)]{fanaroff74} Fanaroff, B.~L., \&
  Riley, J.~M.\ 1974, \mnras, 167, 31P
\bibitem[Fan et al.(2001)]{fan01} Fan, X., Narayanan, V.~K., Lupton,
  R.~H., et al.\ 2001, \aj, 122, 2833
\bibitem[Grimes et al.(2004)]{grimes04} Grimes, J.~A., Rawlings, S.,
  \& Willott, C.~J.\ 2004, \mnras, 349, 503
\bibitem[Jackson \& Wall(1999)]{jackson99} Jackson, C.~A., \& Wall,
  J.~V.\ 1999, \mnras, 304, 160
\bibitem[Hasinger(2004)]{hasinger04} Hasinger, G.\ 2004, Nuclear Physics B
  Proc. Supp., 132, 86
\bibitem[Infante(1994)]{infante94} Infante, L.\ 1994, \aap, 282, 353
\bibitem[Kellermann et al.(1998)]{kellermann98} Kellermann, K. I.,
  Vermuelen, R. C., Zensus, J. A., \& Cohen, M. H. 1998, AJ, 115, 1295
\bibitem[Kellermann et al.(2004)]{kellermann04} Kellermann, K.~I.,
  Lister, M.~L., Homan, D.~C.  et al.,\ 2004, \apj, 609, 539
\bibitem[Lister \& Homan(2005)]{lister05} Lister, M. L., \& Homan,
  D. C. 2005, \aj, 130, 1389
\bibitem[Lister(2003)]{lister03} Lister, M. L. 2003, \apj, 599, 105
\bibitem[Lister \& Marscher(1997)]{lister97} Lister, M.~L., \&
  Marscher, A.~P.\ 1997, \apj, 476, 572
\bibitem[Longair(1966)]{longair66} Longair, M.~S.\ 1966, \mnras, 133,
  421
\bibitem[Longair \& Scheuer(1970)]{longair70} Longair, M. S., \&
  Scheuer, P. A. G. 1970, MNRAS, 151, 45
\bibitem[Osmer \& Smith(1970)]{osmer70} Osmer, P. S., \& Smith,
  M. G. 1980, \apjs, 42, 333
\bibitem[Peacock et al.(1981)]{peacock81} Peacock, J. A., Perryman,
  M. A. C., Longair, M. S., Gunn, J. E., \& Westphal, J. A.\ 1981,
  \mnras, 194, 601
\bibitem[Peebles(1980)]{peebles80} Peebles, P.~J.~E.\ 1980, The
  large-scale structure of the universe, Princeton University Press,
  435
\bibitem[Scheuer(1987)]{scheuer87} Scheuer P.\,A.\,G., 1987,
  Superluminal Radio Sources, Cambridge University Press, Cambridge,
  ed. J.A.~Zensus J., T.~Pearson, p. 104.
\bibitem[Schmidt(1969)]{schmidt69} Schmidt, M. 1969, ApJ, 151, 393
\bibitem[Schmidt et al.(1991)]{schmidt91} Schmidt, M., Schneider,
  D.~P., \& Gunn, J.~E.\ 1991, The Space Distribution of Quasars,
  ed. D.~Crampton, ASP CS-394
\bibitem[Silverman et al.(2005)]{silverman05} Silverman, J.~D., Green,
  P.~J., Barkhouse, W.~A., et al.\ 2005, \apj, 618, 123
\bibitem[Taylor et al.(2001)]{taylor01} Taylor, A.~C., Grainge, K.,
  Jones, M.~E., et al.\ 2001, \mnras, 327, L1
\bibitem[Wall et al.(1980)]{wall80} Wall, J.~V., Pearson, T.~J., \&
  Longair, M.~S.\ 1980, \mnras, 193, 683
\bibitem[Wall et al.(2005)]{wall05} Wall, J.~V., Jackson, C.~A.,
  Shaver, P.~A., Hook, I.~M., \& Kellermann, K.~I.\ 2005, \aap, 434,
  133
\bibitem[Willott et al.(1998)]{willott98} Willott, C.~J., Rawlings,
  S., Blundell, K.~M., \& Lacy, M.\ 1998, \mnras, 300, 625
\bibitem[Willott et al.(2001)]{willott01} Willott, C. J., Rawlings, S.,
  Blundell, K. M., et al. 2001, MNRAS, 322, 536
\bibitem[Wills \& Lynds(1978)]{wills78} Wills, D., \& Lynds, R.\ 1978,
  \apjs, 36, 317
\bibitem[Ueda et al.(2003)]{ueda03} Ueda, Y., Akiyama, M., Ohta, K.,
  \& Miyaji, T.\ 2003, ApJ, 598, 886
\bibitem[Urry \& Padovani(1991)]{urry91} Urry, C.~M., \& 
Padovani, P.\ 1991, \apj, 371, 60
\bibitem[Urry \& Padovani(1995)]{urry95} Urry, C.~M., \& Padovani, P.\
  1995, \pasp, 107, 803
\bibitem[Urry \& Shafer(1984)]{urry84} Urry, C.~M., \& Shafer, 
R.~A.\ 1984, \apj, 280, 569 
\bibitem[Vermeulen \& Cohen(1994)]{vermeulen94} Vermeulen, R.~C., 
\& Cohen, M.~H.\ 1994, \apj, 430, 467
\bibitem[Zensus et al.(2002)]{zensus02} Zensus, J. A., Ros, E.,
    Kellermann, K. I., et al. 2002, AJ, 124, 662

\end{thebibliography}

\end{document}